\documentclass{article}
\usepackage{spconf,amsmath,graphicx}

\usepackage{enumitem}
\setlist{nosep, leftmargin=14pt}

\usepackage{microtype}
\usepackage[square, sort, comma, numbers, compress]{natbib}

\begin{document}
\title{Automated Ventricle Parcellation and Evan's Ratio Computation in~Pre-~and~Post-Surgical~Ventriculomegaly}

\address{%
$^{1}$ Department of Biomedical Engineering, Johns Hopkins School of Medicine, USA\\
     $^{2}$ Department of Electrical and Computer Engineering, Johns Hopkins University, USA\\
     $^{3}$ Laboratory of Behavioral Neuroscience, National Institute on Aging,\\ National Institutes of Health, USA\\
     $^{4}$ Department of Radiology, Case Western Reserve University School of Medicine, USA\\
     $^{5}$ Department of Neurosurgery, Johns~Hopkins~School~of~Medicine, USA}
%
%
\maketitle

\begin{abstract}
Normal pressure hydrocephalus~(NPH) is a brain disorder associated with enlarged ventricles and multiple cognitive and motor symptoms. The degree of ventricular enlargement can be measured using magnetic resonance images~(MRIs) and characterized quantitatively using the Evan's ratio~(ER).  
Automatic computation of ER is desired to avoid the extra time and variations associated with manual measurements on MRI.  Because shunt surgery is often used to treat NPH, it is necessary that this process be robust to image artifacts caused by the shunt and related implants.  In this paper, we propose a 3D regions-of-interest aware~(ROI-aware) network for segmenting the ventricles. 
The method achieves state-of-the-art performance on both pre-surgery MRIs and post-surgery MRIs with artifacts. Based on our segmentation results, we also describe an automated approach to compute ER from these results. 
Experimental results on multiple datasets demonstrate the potential of the proposed method to assist clinicians in the diagnosis and management of NPH.  
\end{abstract}

\section{Introduction}
\begin{figure}[!tb]
\centering
\includegraphics[width=1\columnwidth]{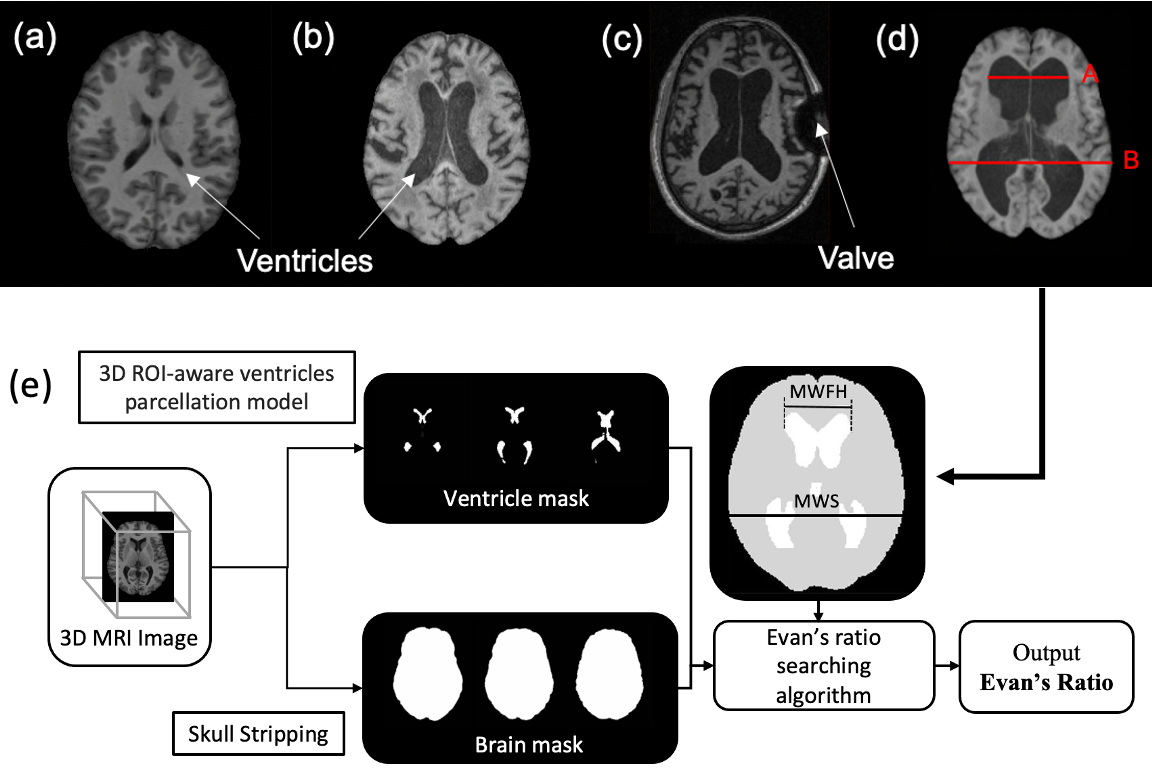} 
\caption{T1-weighted MRIs showing the lateral ventricles of: \textbf{(a)}~a healthy subject, \textbf{(b)}~an NPH subject, and \textbf{(c)}~a post-surgical subject with an MRI artifact.  \textbf{(d)}~Evan's ratio is A/B. \textbf{(e)}~Flowchart of automated Evan's ratio computation, including the measurement of maximum width of frontal horns~(MWFH) and maximum width of inner skull~(MWS).
}
\label{fig::mri_images}
\end{figure}

Normal pressure hydrocephalus~(NPH), presenting as ventriculomegaly, is a chronic disease with symptoms of cognitive impairment, gait dysfunction, and dementia~\cite{shprecher2008normal}. 
Compared with healthy subjects, the ventricles of NPH patients are greatly expanded with excess cerebrospinal fluid~(CSF) causing distortion of the human brain (see Figs.~\ref{fig::mri_images}(a) and (b)). However, unlike Parkinson's, Alzheimer's, and other neurological diseases, the symptoms of NPH are potentially reversable by CSF valve shunting surgery to remove some excess CSF~\cite{stein2006shunts}. 

The diagnosis of NPH remains challenging because the symptoms of NPH overlap with various forms of dementia. 
Evan's ratio~(ER)~\cite{brix2017evans} as illustrated in Fig.~\ref{fig::mri_images}(d), is often used to analyze brain magnetic resonance images~(MRIs) for possible NPH and to monitor the effectiveness of shunt surgery. 
To date, the measurements required for computation of ER have been made manually, which is time-consuming and prone to error. In this paper, we describe an automatic procedure, illustrated in Fig.~\ref{fig::mri_images}(e), which segments and labels the ventricles, measures key dimensions in a normalized space, and outputs the ER.    

FreeSurfer~\cite{fischl2012freesurfer}, RUDOLPH~\cite{carass2017whole}, VParNet~\cite{shao2018dlf, shao2019brain}, and several other methods~\cite{shiee2011ipmi, ellingsen2016spie, atlason2019spie, wang2023spie} provide ventricle segmentations from MRIs.
FreeSurfer is an atlas-based approach for whole brain segmentation; it requires long processing time and may fail on NPH subjects with highly enlarged ventricles or post-surgery MRI artifacts~\cite{lemcke2013safety}. 
Although RUDOLPH is specially designed for subjects with enlarged ventricles, it has a multi-hour run time and often fails on post-surgical subjects. 
VParNet uses a 3D U-net to perform ventricle segmentation in about 2~minutes. It works well on both NPH and healthy subjects, but it also fails when MRI artifacts are present (see Fig.~\ref{fig::visual_result}(c)).

\begin{figure*}[!tb]
\centering
\includegraphics[width=1\textwidth]{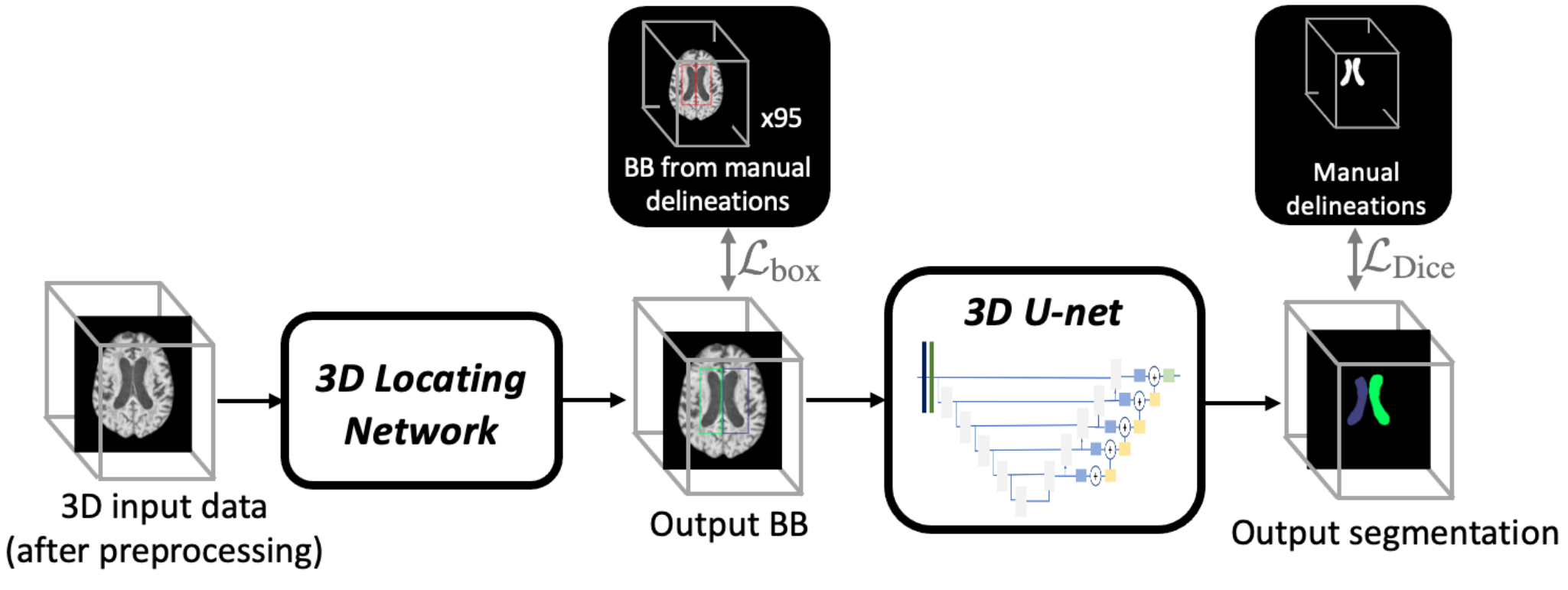}
\caption{Flowchart of our ventricle parcellation. The locating network finds a 3D ROI around the ventricles system, 
then the segmentation network segments the primary ventricle cavities. BBs refer to bounding boxes.}
\label{fig::workflow}
\end{figure*}

In this paper, we propose a novel ventricle segmentation algorithm based on 
localization network and a modified 3D U-net~\cite{cciccek20163d}. 
The method first finds a region-of-interest~(ROI) for the ventricles systems. 
The ROIs are then used to restrict the area of interests of a follow-up segmentation model such that it can focus on the ventricle regions and ignore outer regions that sometimes affected by image artifacts.
We conducted comprehensive experiments on three datasets: one dataset with healthy controls, one dataset of NPH patients, and one dataset of NPH subjects with post-surgery implants (PS-NPH). 
Our method consistently outperformed existing methods, especially on subjects with MRI artifacts. 
Built upon the successful segmentation of ventricles, we then propose an automated ER computation method and validated the agreement of the automated results with manual calculations on 101 subjects. 
The correlation between our manual and automated ER computations was $0.983$, indicating the high reliability of our proposed ER calculation framework. 
Our contributions can be summarized as: 1) First automated computation of Evan's ratio; 2) Validated Evan's ratio on 101 subjects; 3) Improved ventricle segmentation that is robust to post surgical artifacts.

\section{Methods}
An overview of our method is shown in Fig.~\ref{fig::workflow}.
To avoid the effect of MRI artifacts in the image, we use a 3D locating network as the first step; this also reduces the area of interest and thus the complexity of the task in the subsequent steps. The locating network uses the architecture in~\cite{han2020automatic} and is designed to generate four separate 3D bounding boxes, two for the left and right lateral ventricles and one each for the third and fourth ventricles.  We use instance normalization with a small batch size for memory efficient training of the locating network on 3D volumes.
The segmentation network is based on the 3D U-net~\cite{cciccek20163d} with instance normalization and nearest-neighbor interpolation for upsampling.
It takes the ROI-cropped images as input and segments the left lateral~(LLV), right lateral~(RLV), third~(3V), and fourth~(4V) ventricles. 
The cerebral aqueduct is included in the third ventricle label.

\noindent\textbf{3D locating network:} For the locating network, 
all images were rigidly registered to a standard MNI space. The ground truth bounding boxes were obtained from the manual delineations of the four ventricles, where the maximum and minimum coordinates were used as the starting and stopping coordinates of the bounding box. The loss function to train the locating network is given by
\begin{eqnarray*}
    \mathcal{L}_{\text{box}} & = & \frac{1}{N} \sum_{i=1}^{N} s (\hat{x}_{i} - x_{i}), ~~ 
    \text{where } \\
    s(u) & = & \left\lbrace \begin{array}{lcl} 0.5u^{2} & \text{if }~|u| < 1, \\  |u| - 0.5 & \text{otherwise}. \end{array} \right.
\end{eqnarray*}
Here, $\hat{x}_{i}$ is a predicted bounding box coordinate with corresponding ground truth $x_{i}$. There are $N=24$, which corresponds to the two vertices for each of the four desired bounding boxes. 
Our locating network was trained for 500 epochs using the Adam optimization algorithm with a learning rate of $\alpha = 10^{-3}$.

\noindent\textbf{Ventricle parcellation network:} The locating network finds four tight bounding boxes of varying sizes. To accommodate our parcellation network, we expand the bounding boxes symmetrically in all six cardinal directions so that each dimension is a multiple of 32. The segmentation network is trained with the loss
\begin{eqnarray*}
%
\mathcal{L}_{\text{Dice}} & = \left( 1 - \frac{1}{L} \sum_{l = 1}^{L} \frac{\varepsilon + 2\sum_{v} M_{vl} N_{vl}}{\varepsilon + \sum_{v} M_{vl} + \sum_{v} N_{vl}} \right)\!,
\label{eq:dice_loss_box}
\end{eqnarray*}
where $M_{vl}$ is the probability that voxel $v$ has label $l$ generated by the network after a softmax, $N_{vl}$ is the binary value indicating if voxel $v$ should be labeled $l$. 
%
$\varepsilon$ $ = 10^{-3}$ is used to avoid a zero denominator during training. Data augmentation during training includes random left-right flipping, elastic deformation, and rotation. Our network was trained for 150 epochs using the Adam optimizer with a learning rate of $\alpha = 10^{-3}$.

\noindent\textbf{Automated Evan's ratio calculation:} Based on our ventricle segmentation result and using a brain mask from~\cite{roy2017robust}, we automated the Evan's ratio~(ER) calculation as illustrated in Fig.~\ref{fig::mri_images}(f). Recall that all images and corresponding masks are in MNI space. ER is calculated as the maximum width of the frontal horns~(MWFH) from the lateral ventricle masks divided by the maximum width of the inner skull~(MWS). Both the MWFH and MWS are identified as horizontal lines in MNI space, as such we simply search our lateral ventricle masks for the MWFH, and the skull mask for the MWS.

\section{Experiments and Results}
\label{s:expt}
\noindent\textbf{Datasets and Pre-processing:} Magnetization-prepared rapid gradient-echo~(MPRAGE) T1-weighted~(T1-w) images from three cohorts were used to train and validate our proposed method. 
Four ventricle compartments---i.e., the LLV, RLV, 3V, and 4V---were manually delineated. The first dataset contains 50 MRIs of healthy controls from Neuromorphometrics Inc.~(NMM)~\cite{marcus2007open}; from these, 15 images were randomly selected for training, 5 for validation, and the remaining 30 were used for testing. The second dataset contains 95 NPH subjects~\cite{shao2019brain}; of these, 25 images were used for training, 5 for validation, and the remaining 65 for testing. The third cohort contains 6 post-surgical images; these we used exclusively for testing. All images were pre-processed using N4 inhomogeneity correction~\cite{tustison2010tmi}  and rigid registration to MNI space.

\begin{figure}[!tb]
\centering
\includegraphics[width=1\columnwidth]{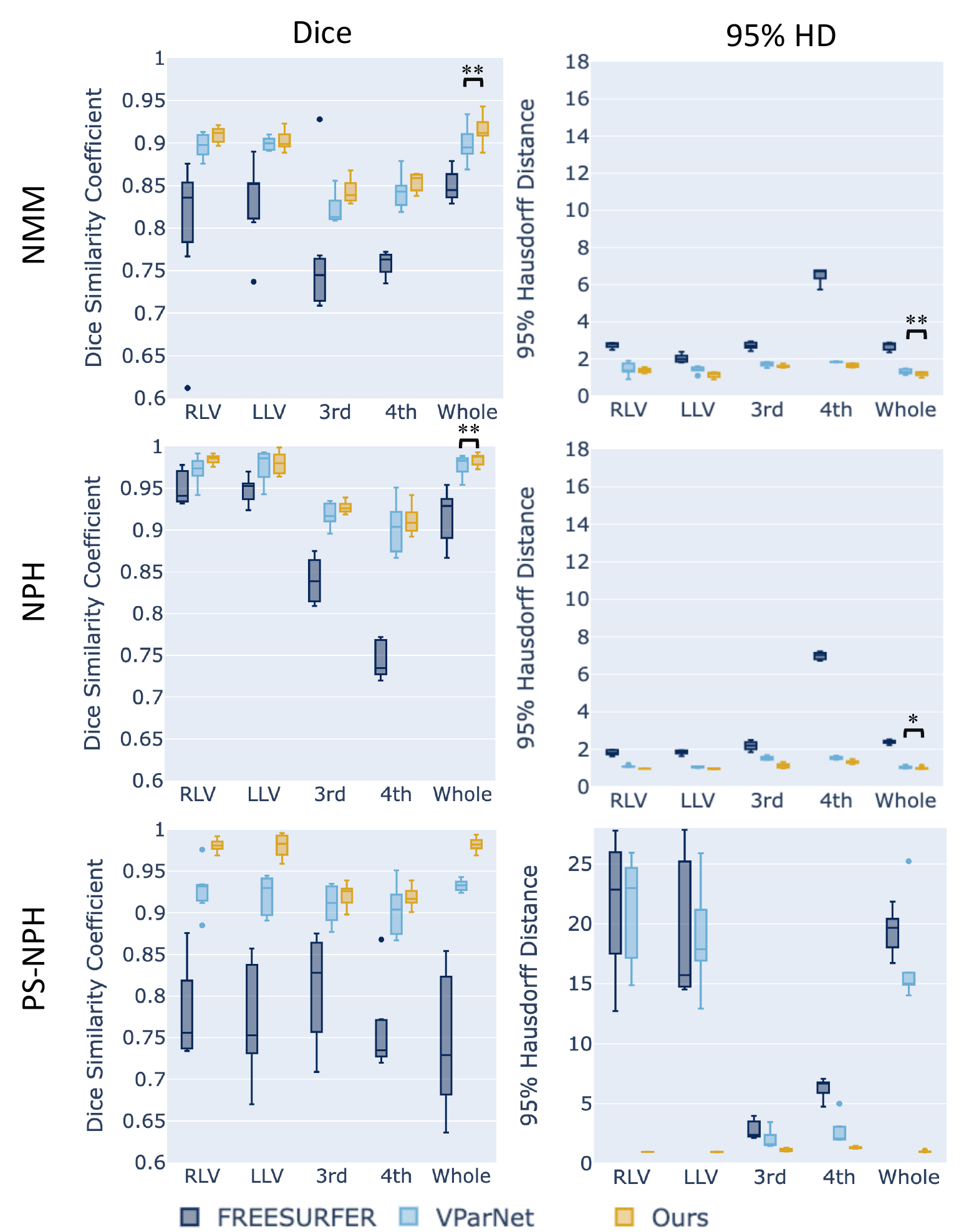} 
\caption{Box plots of DSC and 95\%~HD on NMM, NPH, and PS-NPH test datasets.  Wilcoxon signed-rank test was conducted between VParNet and our method. Our results were significantly better in all datasets (marked by asterisks: $*$ for $p < 0.05$ and $**$ for $p < 0.01$.)} 
\label{fig::dice_hd_result}
\end{figure}

\begin{figure}[!tb]
\centering
\includegraphics[width=0.85\columnwidth]{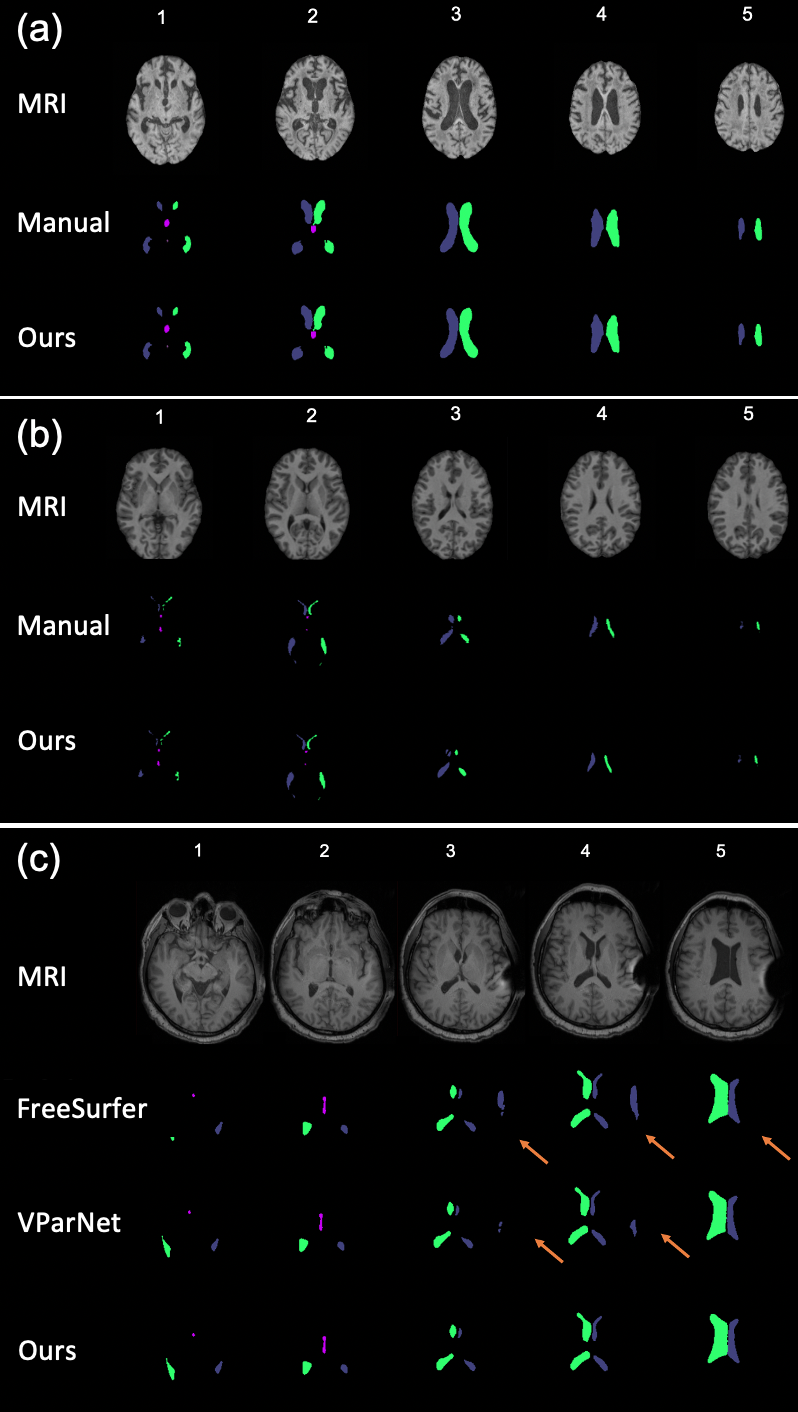} 
\caption{Comparison of three methods for \textbf{(a)}~an NMM, \textbf{(b)}~an NPH, and \textbf{(c)}~a PS-NPH subject; five slices for each subject.} 
\label{fig::visual_result}
\end{figure}

In our first experiment, we compare our proposed method to FreeSurfer~\cite{fischl2012freesurfer} and VParNet~\cite{shao2019brain} on the 30 testing subjects from the NMM cohort, 65 subjects from our NPH cohort, and 6 post-surgery subjects. We computed the Dice similarity coefficient~(DSC) and the 95\%~Hausdorff distance (HD) for these three datasets, and report results in Fig.~\ref{fig::dice_hd_result}.
For all three cohorts, our proposed method is significantly better than VParNet based on  the Wilcoxon signed-rank test (see Fig.~\ref{fig::dice_hd_result} for significance levels). A visualization of the ventricle parcellation produced by our method is shown in Figs.~\ref{fig::visual_result}~(a), (b), and~(c) for the NMM, NPH, and PS-NPH data, respectively. As illustrated in Fig.~\ref{fig::visual_result}~(c), FreeSurfer and VParNet both have failure cases in the area of the post-surgery valve artifact.

\begin{figure}[!tb]
\centering
\includegraphics[width=1\columnwidth]{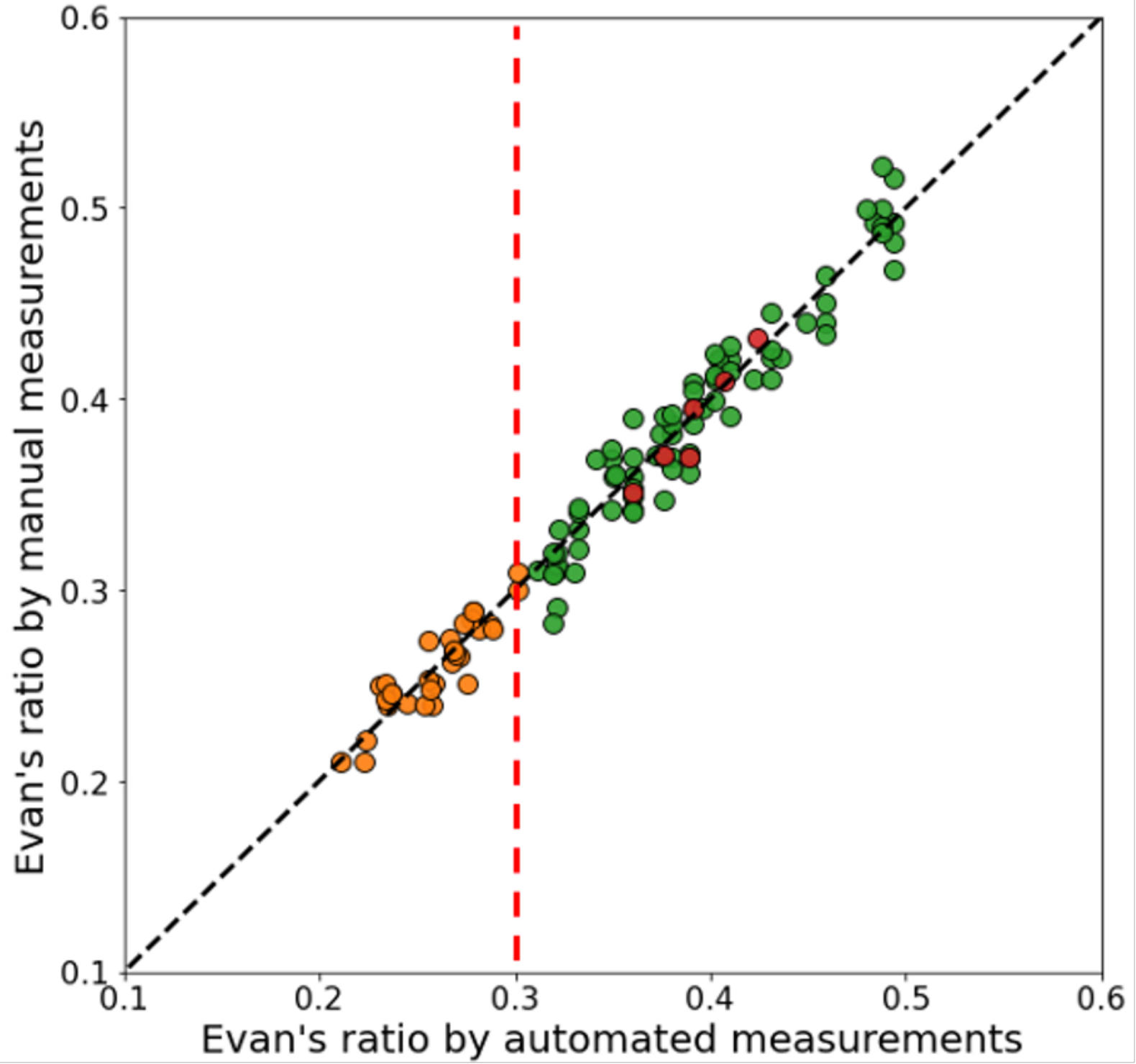} 
\caption{Comparison of Evan's ratio between the automated calculation and the manual measurement.} 
\label{fig::evans_ratios_results}
\end{figure}

In our second experiment, we compare the Evan's ratio~(ER) value from our method with the ER identified by manual measurements on the same testing subjects as in our parcellation experiment (30~NMM subjects, 65~NPH subjects, and 6~PS-NPH subjects). The manual measurements of ER were obtained by averaging the ER determined by two annotators. Figure~\ref{fig::evans_ratios_results} shows a scatter plot of our automatic results and the manual measurements; the correlation coefficient between these two measurements is 0.983. The mean difference between the automated and manual measurements is 0.008 (1.4\%). We note that the automated and manual measurements of ER separated the healthy subjects~(NMM cohort) and NPH subjects~(NPH and PS-NPH cohorts)  at a threshold of $0.3$, which coincides with the threshold used by clinicians for NPH diagnosis~\cite{shprecher2008normal, zhou2021application}.

\section{Conclusions}
In this paper, we proposed a cascade workflow 
to regulate multi-ROIs for robust parcellation of ventricles in subjects with normal and enlarged ventricles, as well as those with post-surgery MRI artifacts. Compared with the current state-of-the-art methods, our method achieves superior results in both qualitative and quantitative evaluations on three datasets. In addition, our proposed method is the only one that handled patients with post-surgery images containing artifacts, demonstrating the robustness of the proposed ROI-aware segmentation. We also presented an automated ER calculation method to assist with diagnosis and monitoring of NPH. The correlation coefficient of ER between our automatic and the manual measurement was 0.983, demonstrating the clinical potential of our method.


\section{Acknowledgments}
\label{sec:acknowledgments}
This work was supported in part by the NIH / NINDS under grant U01-NS122764 (PI: M.G.~Luciano) and in part by the Intramural Research Program of the NIH, National Institute on Aging. Portions of the used data in this study was conducted retrospectively using human subject data made available by Neuromorphometrics Inc. Ethical approval was not required as confirmed by the license attached with this data. The remainder of the data was acquired in line with the principles of the Declaration of Helsinki. Approval was granted by an IRB Committee of the Johns Hopkins School of Medicine with approval IDs CIR00002740~(approved June 6, 2014) and IRB00305245~(approved January 13, 2022).

\bibliographystyle{IEEEtran}
\bibliography{mybib}

\end{document}